# Towards a systematic design of isotropic bulk magnetic metamaterials using the cubic point groups of symmetry

J. D. Baena, L. Jelinek, R. Marqués

Departamento de Electrónica y Eletromagnetismo. Universidad de Sevilla.
Avda. Reina Mercedes s/n, 41012-Sevilla (Spain). Phone: (+34) 954550959
E-mail: juan_dbd@us.es

**Abstract.** *In this paper a systematic approach to the design of bulk isotropic magnetic metamaterials is presented. The role of the symmetries of both the constitutive element and the lattice are analyzed. For this purpose it is assumed that the metamaterial is composed by cubic SRR resonators, arranged in a cubic lattice. The minimum symmetries needed to ensure an isotropic behavior are analyzed, and some particular configurations are proposed. Besides, an equivalent circuit model is proposed for the considered cubic SRR resonators. Experiments are carried out in order to validate the proposed theory. We hope that this analysis will pave the way to the design of bulk metamaterials with strong isotropic magnetic response, including negative permeability and left-handed metamaterials.*

## I. INTRODUCTION

Metamaterials are artificial media exhibiting exotic electromagnetic properties not previously found in Nature. Among them, media showing simultaneously negative electric permittivity and magnetic permeability in some frequency range, or "left-handed" metamaterials, are of particular interest. The striking properties of left-handed metamaterials, including backward-wave propagation, negative refraction, and inverse Cerenkov and Doppler effects were first reported by Veselago [1] in 1968. However, the realistic implementations of left-handed metamaterials came several decades later, as a combination of Split Ring Resonators (SRRs) and metallic wires [2]. SRRs are small planar resonators exhibiting a strong magnetic response, which were proposed in 1999 by J. B. Pendry [3] as suitable "atoms" for the development of negative magnetic permeability metamaterials. One year later, D. R. Smith and co-workers demonstrated the possibility of making up a left-handed medium by periodically combining metallic wires – which provide an effective negative permittivity at microwaves [4] – and SRRs [2]. In subsequent works, other SRR designs were proposed [5-8], in order to reduce electrical size and/or cancel the bi-anisotropic behavior of the original Pendry's design. However, all the aforementioned implementations of negative permeability and left-handed metamaterials are highly anisotropic – or even bi-anisotropic [5] – providing only a uniaxial resonant magnetization; while isotropy is needed for many interesting applications of metamaterials, as for instance the "perfect lens" proposed by Pendry [9].

The aforementioned realizations are, in fact, a combination of two separate systems, one providing the negative magnetic permeability (the SRRs system) and other providing the negative electric permittivity (the wires system). How both sub-systems can be combined in order to obtain a new system whose electromagnetic properties were mainly the superposition of the magnetic and the electric properties of each sub-system is an interesting and controversial issue [10, 11], which is however beyond the scope of this paper. In what follows we will assume that it is possible to find some combination of two isotropic sub-systems, one made of metallic wires (or other elements providing a negative electric permittivity) and the other made of SRRs, whose superposition gives a left-handed metamaterial, and will focus our attention on the design of isotropic systems of SRRs. Actually, since isotropic media with negative magnetic permeability are not found in Nature, an isotropic system of SRRs providing such property in some frequency range will be an interesting metamaterial by itself. These metamaterials could provide the dual of negative electric permittivity media, with similar applications (in imaging [9], for instance). They would be also of interest for magnetic shielding and other practical applications.

A first attempt to design an isotropic magnetic metamaterial was carried out by P. Gay-Balmaz and O. Martin [12], who designed a spherical magnetic resonator – formed by two SRRs crossed in right angle – which is isotropic in two dimensions. This result was later generalized in [13], where a fully isotropic spherical magnetic resonator was proposed. However, from a practical standpoint, it is usually easier to work with cubic designs. A first attempt on such direction was made by C. R. Simovski and co-workers in [14-16], where cubic arrangements of planar SRRs and omega particles were proposed (see Fig. 1(a,b)). If only the magnetic/electric dipole representations of the SRRs and/or Omega particles are considered, these arrangements are invariant under cubic symmetries. However, it has been shown [13, 17] that this invariance is not enough to guarantee an isotropic behavior, since couplings between the planar resonators forming the cubic arrangement can give rise to an anisotropic behavior, even if its dipole representations suggest an isotropic design. The first isotropic metamaterial design fully invariant under the whole group of symmetry of the cube was proposed and simulated in [18]. It is formed by volumetric square SRRs with four gaps, in order to provide 90º rotation symmetries about any of the cube axes. However, this design is unfortunately very difficult to implement in practice, because it cannot be manufactured by using standard photo-etching techniques, as previous SRR designs [2, 3, 5-8, 13-17], and the gaps of the SRR have to be filled with a high relative permittivity dielectric (about one hundred). The idea of using spatial symmetries to design isotropic metamaterials was further developed in [13, 17, 19], leading to the structures depicted in Fig. 1(c,d).

A second group of attempts to design isotropic metamaterials is developed in [20] and [21]. In these works lattices of dielectric and/or paramagnetic spheres with very high refractive index are proposed. If the refractive index of the spheres is high enough, the internal wavelength becomes small with regard to the macroscopic wavelength, and Mie resonances of the spheres can be used to produce the negative effective permittivity and/or permeability. Since the metamaterial "atoms" are spheres, the isotropy is ensured by simply placing them in a cubic lattice. However, practical difficulties to implement such proposals are not easy to overcome. First of all, lossless media with the very high refractive index needed for the spheres are difficult to obtain. Secondly, the system has a very narrow band behaviour [21].

All the previously reported proposals for isotropic magnetic metamaterial design use a "crystal-like" approach. That is, they are based on the homogenization of a system of magnetic resonators which, according to causality laws, exhibit a strong diamagnetic response above resonance. There is, however, another approach widely used in the microwave community which is based on the transmission line analogy to effective media. Initially proposed for two-dimensional metamaterial design [22], it was recently generalized to three-dimensional isotropic structures [23-26]. The main advantage of this approach is its broadband operation, since no resonators are necessary for the design. However, it also presents disadvantages with regard to "crystal-like" approaches. The transmission line approach to metamaterials does not seem to be applicable beyond the microwave range, whereas a significant magnetic response of the SRR has been shown in the terahertz range and beyond [27, 28]. In addition, the coupling to free space of the reported transmission line metamaterials seems to be difficult, and sometimes needs of an additional specific interface (e.g. an antenna array [25]), whereas this coupling is direct in crystal-like metamaterials.

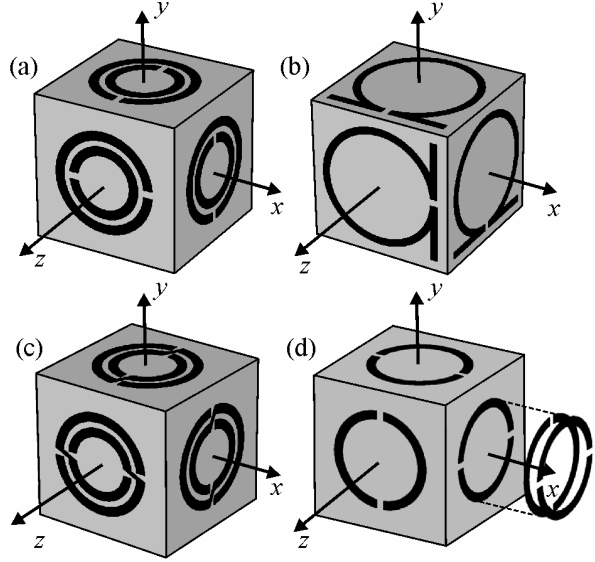

**Fig. 1:** Cubic constitutive elements for isotropic metamaterial design. The cubes (a) and (b) were studied in Refs. [14-16]. Their hidden faces are arranged in such a way that the cube satisfy the central symmetry to avoid magnetoelectric coupling. The cubes (c) and (d) were proposed in Ref. [13] as truly 3D isotropic cubic resonators.

Finally, regarding isotropic left-handed metamaterial design, it should be mentioned that some recent proposals based on random arrangements of chiral particles [29, 30] have the advantage of providing simultaneously both electric and magnetic negative polarizability. This approach can be straightforwardly extended to the design of SRR magnetic metamaterials, by simply considering random arrangements of such elements. There is however a major difficulty with this approach: the constitutive elements in a random composite have to be very small in comparison with the macroscopic wavelength to show a true statistical behaviour, but it is not easy to design a SRR much smaller than one tenth of the wavelength. Due to this fact, periodic arrangements will be considered in what follows.

The main aim of this paper is to present a systematic approach to the design of metamaterial structures based on periodic arrangements of SRRs. The first section is focused on the spatial symmetries which are necessary to ensure an isotropic behavior in the metamaterial. Cubic arrangements of SRRs placed on cubic lattices are considered, and the minimum symmetry requirements for both, the individual resonators and the lattices are investigated. The second section is devoted to a deeper analysis of the isotropic cubic SRR resonators forming the basis of the crystal structure. An equivalent circuit model for such cubic SRR resonators is developed and applied to some specific examples. The third section is focused on the experimental verification of the analysis developed in the previous ones. Finally, the main conclusions of the work are presented.

## II. ROLE OF CUBIC SYMMETRIES

Let us assume that constitutive elements and the unit cell of the material are much smaller than the operating wavelength. In such a case, the interaction of electromagnetic field with the material is described by means of constitutive relations. Besides, the material is supposed to be linear, so the most general way to express those relations between electromagnetic intensities and electromagnetic flux densities is [31]

$$\begin{aligned} \mathbf{D} &= \boldsymbol{\varepsilon} \cdot \mathbf{E} + \boldsymbol{\xi} \cdot \mathbf{H} \\ \mathbf{B} &= \boldsymbol{\zeta} \cdot \mathbf{E} + \boldsymbol{\mu} \cdot \mathbf{H} \end{aligned} \quad (1)$$

where $\boldsymbol{\varepsilon}, \boldsymbol{\mu}$, are second rank constitutive tensors and $\boldsymbol{\xi}, \boldsymbol{\zeta}$ are second rank constitutive pseudo-tensors. In order to get a macroscopic isotropic behaviour, all constitutive tensors and pseudo-tensors $\boldsymbol{\varepsilon}, \boldsymbol{\mu}, \boldsymbol{\xi}$, and $\boldsymbol{\zeta}$ must become scalars or pseudo-scalars.

Let us now address the problem of forcing the tensors/pseudo-tensors in (1) to be scalars/pseudo-scalars for the specific case of a periodic structure. It is well known [32, 33] that there are 32 symmetry point groups for periodic crystals which can be classified in 7 crystallographic systems. It is also known that the cubic system is the only one that forces to any second rank tensor (or pseudo-tensor) to be a scalar (or a pseudo-scalar) [33]. Since any material satisfying the linear constitutive relations (1) and being invariant under the cubic symmetries exhibits an isotropic macroscopic behavior, this section will be focused on the analysis of such cubic symmetries. It is clear that any structure invariant under all the symmetry transformations of the cube must be isotropic, as it was already proposed by Koschny et al. [18]. Furthermore, the full symmetry group of the cube contains four different subgroups also belonging to the cubic system and, thus, providing an isotropic macroscopic behavior. Since a less symmetric design is subjected to less structural constraints, it may be guessed that using these subgroups – instead of the whole symmetry group of the cube – may have practical advantages. Keeping this in mind, first we will give a short overview on the five cubic point groups. Next, we shall connect these point groups with some real structures made of planar resonators commonly used in metamaterials. This will be done in two parts: the study of the symmetries of the constitutive element, or the *basis*, and the analysis of the suitable periodic arrangements, or the *lattice*. At the end of the section some practical isotropic structures will be specifically analyzed.

### a. Cubic point groups

The five cubic point groups are schematically represented in Fig. 2. Following Schöenflies' notation and ordering by degree of symmetry, these groups and their generators are:

- $T = <\{\mathbf{1}, \mathbf{4}_x \cdot \mathbf{4}_y, \mathbf{4}_y \cdot \mathbf{4}_x\}> =$ proper rotations of the regular tetrahedron (12 operations);
- $T_h = <\{\mathbf{1}, \mathbf{-1}, \mathbf{4}_x \cdot \mathbf{4}_y, \mathbf{4}_y \cdot \mathbf{4}_x\}> = T$ expanded by the inversion (24 operations);
- $T_d = <\{\mathbf{1}, \mathbf{-2}_x, \mathbf{4}_x \cdot \mathbf{4}_y, \mathbf{4}_y \cdot \mathbf{4}_x \}> =$ proper and improper rotations of the regular tetrahedron (24 operations);
- $O = <\{\mathbf{1}, \mathbf{4}_x, \mathbf{4}_y\}> =$ proper rotations of the cube (24 operations);
- $O_h = <\{\mathbf{1}, \mathbf{-1}, \mathbf{4}_x, \mathbf{4}_y\}> =$ full symmetry group of the cube (48 operations).

We have used a widely used notation for symmetry transformations, **1** being the identity operator, **-1** the inversion, $\mathbf{n}_p$ a n-fold rotation axis about the *p*-axis, and $\mathbf{-n}_p$ the *n*-fold axis about the *p*-axis followed by the inversion. For example, the operator **-2**$_x$ is the rotation through 180º about the *x*-axis followed by an inversion.

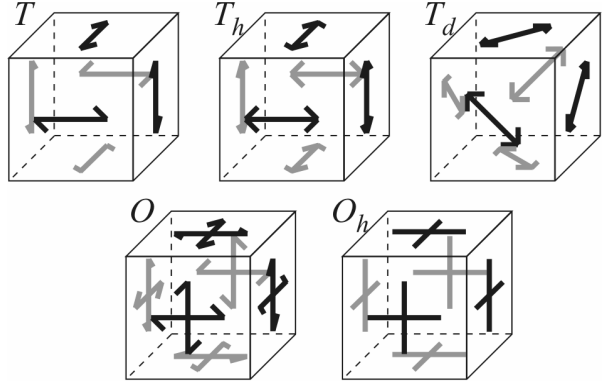

**Fig. 2:** Objects with the symmetries of the five cubic point groups.

### b. Cubic basis

In order to simplify the problem, the symmetries can be separately imposed on the basis and the lattice of the structure. For the sake of simple fabrication, we will assume that the basis is formed by six planar resonators placed over the faces of an inert rigid cube, as in Fig. 1. If the crystal was diluted enough, then the coupling between two neighboring cubes would be much weaker than the coupling between the six SRRs of the same cube and thus each cube could be seen as a single cubic resonator (CR) electromagnetically coupled to others. Such consideration implies that the interaction between the CRs forming the material can be described by dipole-dipole interactions, higher order multipole interactions being negligible. In such approximation all the CRs are properly described by second rank polarizability tensors connecting the external field, $\mathbf{E}^{ext}$ and $\mathbf{B}^{ext}$, with the dipolar moments, $\mathbf{p}$ and $\mathbf{m}$, induced in the CRs [31, 34]:

$$\begin{aligned}\mathbf{p} &= \boldsymbol{\alpha}_{ee} \cdot \mathbf{E}^{ext} + \boldsymbol{\alpha}_{em} \cdot \mathbf{B}^{ext} \\ \mathbf{m} &= \boldsymbol{\alpha}_{mm} \cdot \mathbf{B}^{ext} - \boldsymbol{\alpha}_{em}^{t} \cdot \mathbf{E}^{ext}\end{aligned} \quad , \quad (2)$$

where $\alpha_{ee}$, $\alpha_{mm}$, and $\alpha_{em}$ are the electric, magnetic, and magneto-electric polarizabilities, and the superscript *t* means transpose operation. The constitutive tensors in (1) can be derived from these polarizabilities and from the lattice structure by applying an homogenization technique.

In what follows different kinds of CRs will be named by its cubic group symmetry followed by the acronym CR (group-CR). In order to design an isotropic CR, we have to find suitable planar resonators and place them correctly over the cube so as to fulfill the necessary symmetries. Obviously, the planar resonators have to be invariant under certain symmetry transformations of the square. To classify all different possibilities, a list of the symmetry subgroups of the square is shown in Table I, as well as their geometrical representations, and some examples of planar resonators commonly used in metamaterial design and obeying these symmetries. This table also provides a systematic terminology for planar resonators by using the symbol of the symmetry group followed by the term SRR (group-SRR). In what follows, we will use the term SRR in a general sense covering any type of geometry derived from the SRR and the Omega particle.

By direct inspection on Fig. 2, it can be seen that any of the five cubic point groups contains three 2-fold rotation axes (180º rotations) parallel to the edges of the cube. Thus, only resonators belonging to the last five rows of Table I are appropriate for designing isotropic CRs. At this point, it may be worth to mention that Pendry's SRRs [3], as well as Omega particles [35], are not appropriate for such purpose, because they correspond to the $C_1$-SRR and $D_1$-SRR typologies. In summary, in order to get an isotropic CR, we have to choose six identical SRRs pertaining to the classes $C_2$-, $D_2$-, $C_4$-, or $D_4$-SRR, and arrange them according to one of the cubic point groups $T$, $T_d$, $T_h$, $O$, or $O_h$ shown in Fig. 2.

**Table I.** Classification of SRR types based on the symmetry subgroups of the square. The second column shows the Schöenflies' notation and the generator of groups. The symbols of transformations are: **1** = identity; **4** = 90º rotation; **2** = 180º rotation; **-4** = -90º rotation; $m_x$, $m_y$ = line reflections respect to the $x$ and $y$ axis, respectively; $m_{x,y}$, $m_{x,-y}$ = line reflections respect to both diagonals of the square. Each group is schematically represented by the objects in the second column which can be replaced by the planar resonators shown in the third column.

| SRR types | Symmetry subgroups of the square | Geometrical representation | Examples of resonators |
|---|---|---|---|
| $C_1$-SRR | $C_1 = \{\mathbf{1}\}$ | 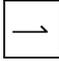 | 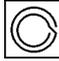 |
| $D_1$-SRR | $D_{1x} = \{\mathbf{1}, \mathbf{m}_x\}$ | 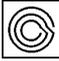 | 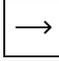 |
| | $D_{1y} = \{\mathbf{1}, \mathbf{m}_y\}$ | 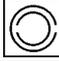 | 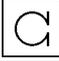 |
| | $D_{1x,y} = \{\mathbf{1}, \mathbf{m}_{x,y}\}$ | 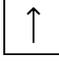 | 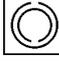 |
| | $D_{1x,-y} = \{\mathbf{1}, \mathbf{m}_{x,-y}\}$ | 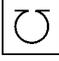 | 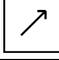 |
| $C_2$-SRR | $C_2 = \{\mathbf{1}, \mathbf{2}\}$ | 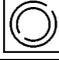 | 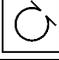 |
| $D_2$-SRR | $D_{2x} = D_{2y} = \{\mathbf{1}, \mathbf{m}_x, \mathbf{m}_y, \mathbf{2}\}$ | 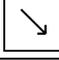 | 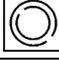 |
| | $D_{2xy} = D_{2x\bar{y}} = \{\mathbf{1}, \mathbf{m}_{x,y}, \mathbf{m}_{x,-y}, \mathbf{2}\}$ | 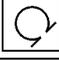 | 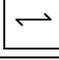 |
| $C_4$-SRR | $C_4 = \{\mathbf{1}, \mathbf{4}, \mathbf{2}, \mathbf{-4}\}$ | 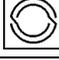 | 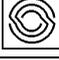 |
| $D_4$-SRR | $D_4 = \{\mathbf{1}, \mathbf{4}, \mathbf{2}, \mathbf{-4}, \mathbf{m}_x, \mathbf{m}_y, \mathbf{m}_{x,y}, \mathbf{m}_{x,-y}\}$ | 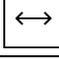 | 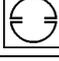 |

Although all five cubic point groups mentioned above are equally useful to achieve isotropic CRs, a specific choice may strongly affect the properties of an isotropic metamaterial. For instance, using isotropic CRs of low symmetry may be quite advantageous since the electrical size of the CRs can be made smaller. This fact can be justified in terms of the $LC$ circuit models for the SRRs [5-8], because the effective capacitance of low symmetry SRRs are usually higher than those of high symmetry SRRs [8], thus providing a smaller resonance frequency. Following these considerations, the best choice of basis would be a $T$-CR made of six planar resonators of the $C_2$-SRR type. A good candidate among all possibilities is the cube shown in Fig. 1(c) made of six non-bianisotropic SRRs (NB-SRRs) [8, 36], a configuration already proposed in Refs. [13, 17]. Furthermore, it was shown in Ref. [13] that this configuration shows a bi-isotropic behavior, due to the lack of inversion symmetry due to the lack of inversion symmetry of the cubic arrangement. However, sometimes an effective isotropic medium without bi-anisotropy ($\xi, \zeta = 0$) is desired. Since $\xi$ and $\zeta$ are pseudo-tensors, the invariance of the CR under inversion is required in order to avoid such property. In this case, the lowest symmetry group is the $T_h$ group. A CR invariant under the last group of symmetry can be made by using planar resonators of the type $D_2$-SRR as, for instance, the symmetric-SRR [37] or the modified double-slit BC-SRR shown in Fig. 1(d) [13]. However, as it will be shown in the following, such symmetry requirements can be relaxed if the lattice symmetries are properly chosen.

### c. Cubic lattices

Above findings give precise instructions for choosing suitable geometries for isotropic metamaterial constitutive elements. The next step is to create an isotropic metamaterial with these elements. The cubic shape of the considered constitutive elements suggest that the best periodical arrangements are the simple cubic (*sc*), body centered cubic (*bcc*), and face centered cubic (*fcc*) lattices shown in Fig. 3. All these lattices obey the full symmetry group of the cube, $O_h$. Therefore, the whole metamaterial (lattice plus basis) retains the cubic point group symmetries and the macroscopic isotropic behavior.

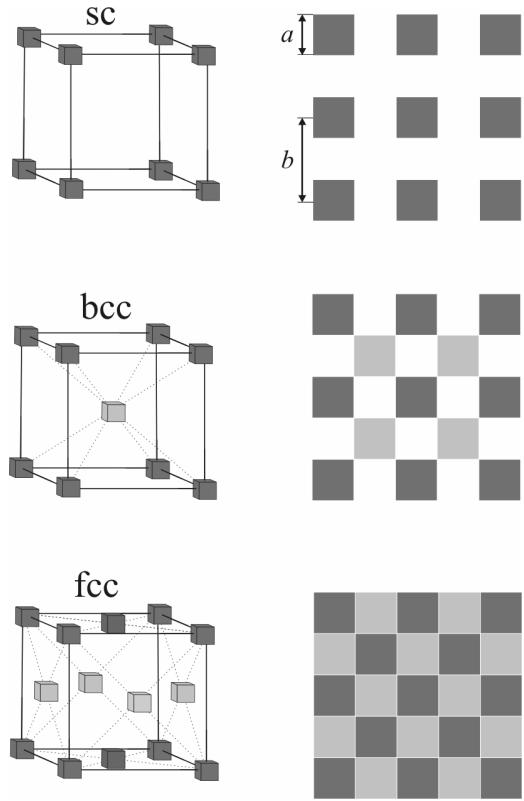

**Fig. 3:** The cubic Bravais's lattices. Their top views are also depicted for the particular case of $b = 2a$. Black and grey small cubes represent cubic resonators on successive planes.

Although all previously mentioned lattices can provide isotropic metamaterials, it is convenient to look deeply into the possible structures, because some particular choices may offer interesting advantages. Regarding Fig. 3, $a$ is the edge size of the CR and $b$ is the edge size of the cubic unit cell. In order to describe CR interactions as dipole-dipole interactions $b$ must be chosen much larger than $a$, so that the metamaterial properties can be deduced from (2) and the appropriate homogenization procedure. However, usually we are also interested in a high density of dipoles in order to get a strong electromagnetic response. Therefore $b$ should be as small as possible. But decreasing $b$ may lad to a failure of the aforementioned homogenization procedure. However, in any case, the combination of a basis and a lattice with the appropriate symmetries will provide an isotropic metamaterial, regardless of the homogenization procedure. Finally, there are some practical limitations to the values that $a$ and $b$ can reach as, for instance, the obvious inequality $b \geq a$, derived from the fact that CRs are supposed to be impenetrable.

Additional limitations appear for each specific structure. In case of an *sc* lattice with $T$-, $T_d$-, or $O$-CRs, the lack of inversion symmetry implies that opposite sides of a CR are not oriented in the same way. Thus, the constrain $b > a$ is necessary in order to avoid a mutual short-circuit between the SRRs of neighboring CRs. To allow the minimum distance $b = a$, the non centro-symmetric CRs in the *sc* lattice must be replaced by $T_h$- or $O_h$-CRs, so that the SRRs on contacting sides of neighboring CRs exactly overlap. In case of a *bcc* lattice, the contact between corners implies that the inequality $b \geq a$ must be fulfilled for any type of CR. Finally, for the *fcc* lattice, the contact between edges of neighboring CRs establishes the harder condition $b \geq 2a$.

The particular case of an *fcc* lattice with the minimum cell size, $b = 2a$, deserves an specific analysis. When $T_h$- or $O_h$-CRs are used as the basis of the *fcc* lattice, the structure turns into an *sc* lattice with the highest possible compactness, i.e. $b = a$, because the *holes* between each eight neighboring CRs have the same shape than the CRs forming the basis. The case of an *fcc* lattice with a $T$-, $T_d$-, or $O$-CR basis is still more special and interesting, because each *hole* exactly corresponds with the inversion of the CRs of the basis. Therefore, although the basis of the structure is not invariant under inversion, the *fcc* structure is brought into coincidence with itself by inversion centered at the center of a CR, followed by a translation of length $a$ through any of the cube axes. Since the wavelength of the signal illuminating the structure is supposed to be much larger than $a$, the system can be consider as macroscopically invariant under inversion and, therefore, any bi-isotropic behavior must disappear. Thus, we conclude that a very interesting choice in order to obtain an isotropic metamaterial is the *fcc* lattice with $b = 2a$ and with a basis formed by $T$-CRs (example in Fig.1(c)), because of its high compactness, non-bi-isotropic macroscopic behavior, and low degree of symmetry. It is worth to recall here that $T$-CRs have the lowest symmetry among all the possibilities shown in Fig. 2, which helps to reduce the electrical size of the unit cell, as it was explained above.

## III.     RESONANCES AND POLARIZABILITIES OF CRs

Until now, only the symmetry of CRs and cubic lattices useful for isotropic periodic metamaterials were analyzed. However, in order to have a complete characterization of the metamaterial, polarizabilities and couplings between individual SRRs must be considered. In dilute crystals the approach of weak coupling between CRs, but strong coupling between the SRRs of each CR is valid. Then, the metamaterial characterization involves two separate problems: obtaining the polarizability tensors in equation (2) for a single CR, and applying the appropriate homogenization procedure to obtain the constitutive parameters for the whole structure. For dense packages, the aforementioned approach is not valid, since couplings between SRRs of different CRs can be stronger than SRR couplings inside each individual CR. However, even in these cases, the analysis of the isolated CR resonances and polarizabilities still provides useful information on the behavior of the metamaterial. For instance, it allows to elucidate if the coupling between SRRs in a practical low symmetry CR can be neglected or not. In case they could be neglected, all the analysis in Section II(b) would become irrelevant, because the SRRs could be substituted by its equivalent dipoles (as it was assumed in [14-16]), without more considerations on the SRR structure. Therefore, the analysis in this section is necessary in order to justify the practical relevance of the analysis developed in Section II. Further, in Sec. IV, an experimental validation of this analysis will be provided.

Let us assume that the CR size is much smaller than the operating wavelength. Thus, an *RLC* circuit model is valid for describing the behavior of single Pendry's SRRs [3], as well as for any type of modified SRRs [5-8] or Omega particles [38]. Then, a CR can be seen as a set of six *RLC* circuits magnetically coupled through its mutual inductances, with one loop of current per each SRR. The positive directions of the electric currents on each loop are arbitrarily defined in Fig. 4. The relation between currents and electromotive forces exciting the CR can be written as

$$\mathbf{Z} \cdot \mathbf{I} = \mathbf{F}, \quad (3)$$

where $\mathbf{Z}$ is a 6 by 6 square impedance matrix, $\mathbf{I}$ is a column matrix whose i-*th* component is the current flowing over the *i-th* SRR, and $\mathbf{F}$ is a column matrix whose i-*th* component is the external electromotive force acting on the *i*-th SRR. The diagonal components of the impedance matrix are the self-impedances of each SRR, i.e. $Z_{ii} = R + j\omega L + 1/(j\omega C)$, with *R*, *L*, and *C* being the resistance, self-inductance, and self-capacitance of a single SRR [8]. The non-diagonal components $Z_{ij}$ are the mutual impedances between the *i*-th and *j*-th SRRs: $Z_{ij} = j\omega M_{ij}$. From the reciprocity theorem [39] we know that the impedance matrix must be symmetric, i.e. $Z_{ij} = Z_{ji}$. This reduces the number of independent elements of $\mathbf{Z}$ to 21. This number can be further reduced by applying the geometrical symmetries of the CR, as shown in the next paragraph.

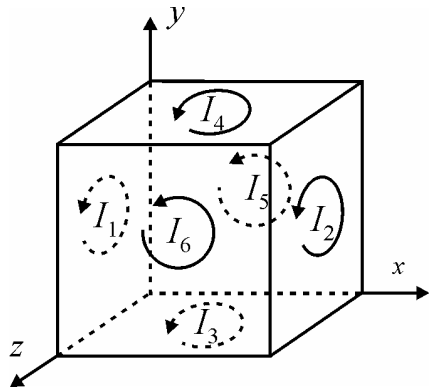

**Fig. 4:** Definition of the sign of currents in the circuit model for a 3D cubic magnetic resonator.

The application of any symmetry operation changes the components of **I** according to the rule **I'** = **S** · **I**, where **S** is the corresponding operator of symmetry. It is well-known that any symmetry operation **S** of the cubic point groups can be expressed as some combination of the three orthogonal 4-fold rotations and the inversion, whose matrix representations, in the six-dimensional space defined by **I**, are:

$$\mathbf{4}_x \equiv \begin{bmatrix} \begin{array}{c|c|c} \begin{matrix} 1 & 0 \\ 0 & 1 \end{matrix} & 0 & 0 \\ \hline 0 & 0 & \begin{matrix} 0 & -1 \\ -1 & 0 \end{matrix} \\ \hline 0 & \begin{matrix} 1 & 0 \\ 0 & 1 \end{matrix} & 0 \end{array} \end{bmatrix} ; \quad \mathbf{4}_y \equiv \begin{bmatrix} \begin{array}{c|c|c} 0 & 0 & \begin{matrix} 1 & 0 \\ 0 & 1 \end{matrix} \\ \hline 0 & \begin{matrix} 1 & 0 \\ 0 & 1 \end{matrix} & 0 \\ \hline \begin{matrix} 0 & -1 \\ -1 & 0 \end{matrix} & 0 & 0 \end{array} \end{bmatrix} .$$

$$\mathbf{4}_z \equiv \begin{bmatrix} \begin{array}{c|c|c} 0 & \begin{matrix} 0 & -1 \\ -1 & 0 \end{matrix} & 0 \\ \hline \begin{matrix} 1 & 0 \\ 0 & 1 \end{matrix} & 0 & 0 \\ \hline 0 & 0 & \begin{matrix} 1 & 0 \\ 0 & 1 \end{matrix} \end{array} \end{bmatrix} ; \quad -\mathbf{1} \equiv \begin{bmatrix} \begin{array}{c|c|c} \begin{matrix} 0 & 1 \\ 1 & 0 \end{matrix} & 0 & 0 \\ \hline 0 & \begin{matrix} 0 & 1 \\ 1 & 0 \end{matrix} & 0 \\ \hline 0 & 0 & \begin{matrix} 0 & 1 \\ 1 & 0 \end{matrix} \end{array} \end{bmatrix}$$

(4)

They are unitary matrices with the well known property $\mathbf{S}^{-1} = \mathbf{S}^t$. It can be straightforwardly demonstrated that **F** follows the same rule of transformation: **F'** = **S** · **F**. Therefore, both **I** and **F** can be considered as vectors. In what follows, **I** and **F** will be called the "current" and the "excitation" vectors, respectively. Therefore, the impedance matrix **Z** is a second rank tensor, following the transformation rule **Z'** = **S** · **Z** · **S**$^t$. If the CR remains invariant by the transformation **S**, then

$$\mathbf{Z} = \mathbf{S} \cdot \mathbf{Z} \cdot \mathbf{S}^t .$$ (5)

This equation gives some relations between the components of **Z**, which can reduce the number of independent components of **Z**.

Although the current vector **I** can be directly solved by multiplying both sides of (3) by $\mathbf{Z}^{-1}$, in order to identify the different resonances of the CR it is convenient to expand the solution in terms of the eigenvectors of **Z**. The eigenvalue problem corresponding to (3) is

$$\mathbf{Z} \cdot \mathbf{v}_i = z_i \mathbf{v}_i ,$$ (6)

where $z_i$ are the eigenvalues, $\mathbf{v}_i$ the eigenvectors, and the index $i = 1 \ldots 6$. The impedance matrix **Z** can be expanded in a sum of two terms as

$$Z_{ij} = \left( R + j\omega L + \frac{1}{j\omega C} \right) \delta_{ij} + Z_{ij}(1 - \delta_{ij}) ,$$ (7)

where $\delta_{ij}$ is the Kronecker's delta. The first term is the self-impedance of a single SRR multiplied by the identity, while the second term is the symmetric matrix of mutual impedances. These mutual impedances are purely imaginary numbers given by $Z_{ij} = j\omega M_{ij}$, where $M_{ij}$ are the mutual inductances between two SRRs. Thus, the second term in (7) is a purely imaginary symmetrical matrix. Therefore, its eigenvectors can be chosen in such a way that they form a complete and orthogonal basis that diagonalizes this matrix. Furthermore, since the first summand in (7) is actually an scalar, the eigenvectors of $Z_{ij}$ are actually the same as those of $Z_{ij}(1 - \delta_{ij})$. Therefore, the eigenvectors of **Z** can be chosen in such a way that they form an orthogonal basis for the considered 6-dimensional space.

Thus, the current and excitation vectors can be expanded as a summation of such eigenvectors:

$$\mathbf{I} = \sum_i (\mathbf{I} \cdot \mathbf{v}_i) \mathbf{v}_i, \quad \mathbf{F} = \sum_i (\mathbf{F} \cdot \mathbf{v}_i) \mathbf{v}_i. \qquad (8)$$

By substituting both expressions into (3) and applying (6), we get

$$\mathbf{I} = \sum_i \frac{\mathbf{F} \cdot \mathbf{v}_i}{z_i} \mathbf{v}_i. \qquad (9)$$

Therefore, both $\mathbf{F}$ and $\mathbf{I}$ can be expanded in a set of orthogonal modes having mutually proportional excitation and current vectors.

From (6) we can also obtain information about the structure of the eigenvalues $z_i$. These eigenvalues must have the form

$$z_i(\omega) = R + j\omega L + \frac{1}{j\omega C} + z_{c,i}(\omega) = j\omega L \left(1 - \frac{\omega_0^2}{\omega^2} + \frac{R}{j\omega L} + \frac{z_{c,i}(\omega)}{j\omega L}\right), \qquad (10)$$

where $\omega_0$ is the resonance frequency an isolated SRR ($\omega_0^2 = 1/LC$), and $z_{c,i}(\omega)$ the eigenvalues of the second summand in (7), which are related with the coupling between SRRs. It can be seen in (9) that the $i$-th mode resonates when its eigenvalue approaches zero ($z_i \approx 0$). Therefore, the frequency of resonance of the $i$-th mode is given by the relation $z_i(\omega_{0,i}) \approx 0$. If losses and couplings between SRRs are not too strong ($R, z_{c,i} \ll j\omega L$) the frequencies of resonance of the CR can be approximated as

$$\omega_{0,i} \approx \omega_0 - \frac{z_{c,i}(\omega_0)}{2 jL}. \qquad (11)$$

In what follows we will apply this method to the determination of the resonances and polarizabilities of two CRs made from two well known SRRs: Pendry's SRR [3] and NB-SRR [8].

### a. Analysis of an anisotropic cube

Let us now consider the CR shown in Fig. 1(a), made of Pendry's SRRs. In this section we are going to get some analytical approximation for its resonances and polarizabilities. Note that the cube possesses inversion symmetry and, thus, magnetoelectric coupling is forbidden, so that $\alpha_{em} = 0$ in (2). It can be also seen by inspection that the considered CR it is invariant under the rotation $\mathbf{4}_y \cdot \mathbf{4}_x$. By applying this spatial symmetry the impedance matrix is reduced to the form

$$\mathbf{Z} = \begin{bmatrix} Z_{11} & Z_{12} & Z_{13} & Z_{14} & -Z_{14} & -Z_{13} \\ Z_{12} & Z_{11} & Z_{14} & Z_{13} & -Z_{13} & -Z_{14} \\ Z_{13} & Z_{14} & Z_{11} & Z_{12} & -Z_{14} & -Z_{13} \\ Z_{14} & Z_{13} & Z_{12} & Z_{11} & -Z_{13} & -Z_{14} \\ -Z_{14} & -Z_{13} & -Z_{14} & -Z_{13} & Z_{11} & Z_{12} \\ -Z_{13} & -Z_{14} & -Z_{13} & -Z_{14} & Z_{12} & Z_{11} \end{bmatrix}, \qquad (12)$$

where there are only four independent components. The corresponding eigenvalues and its orthonormal eigenvectors are shown in Table II. It is worth to note that the eigenvectors can be classified in even and odd modes: for even (odd) modes the currents $I_{2n-1}$ and $I_{2n}$ are parallel (anti-parallel).

**Table II.** Eigenvalues and a complete set of orthonormal eigenvectors of the impedance matrix (4) corresponding to non-isotropic cubic resonators with symmetries **-1** and $\mathbf{4_y \cdot 4_x}$, as for instance the structures shown in Fig. 1(a) and Fig. 1(b).

|  | Eigenvalues, $z_i$ | Eigenvectors, $\mathbf{v}_i$ |
|---|---|---|
| Even modes | $Z_{11} + Z_{12} - Z_{13} - Z_{14}$ | $\frac{1}{2}(-1,-1,1,1,0,0)$ <br> $\frac{1}{2\sqrt{3}}(1,1,1,1,2,2)$ |
|  | $Z_{11} + Z_{12} + 2Z_{13} + 2Z_{14}$ | $\frac{1}{\sqrt{6}}(-1,-1,-1,-1,1,1)$ |
| Odd modes | $Z_{11} - Z_{12} - Z_{13} + Z_{14}$ | $\frac{1}{2\sqrt{3}}(1,-1,-2,2,1,-1)$ <br> $\frac{1}{2}(1,-1,0,0,-1,1)$ |
|  | $Z_{11} - Z_{12} + 2Z_{13} - 2Z_{14}$ | $\frac{1}{\sqrt{6}}(-1,1,-1,1,-1,1)$ |

Once the eigenvalue problem is solved, the next step is to write an explicit expression for the excitation vector **F** and introduce this expression in (9), in order to get the currents over the SRRs. To begin with, we will assume that the CR is excited by a homogeneous external magnetic field $\mathbf{B}^{ext} = (B_x^{ext}, B_y^{ext}, B_z^{ext})$, and there is not any external electric field. Then, the excitation vector is written as

$$\mathbf{F}_m = -j\omega A\, (B_x^{ext}, B_x^{ext}, B_y^{ext}, B_y^{ext}, B_z^{ext}, B_z^{ext}), \tag{13}$$

where $A$ is the effective area of the SRR. By introducing (13) into (9), the current vector **I** is calculated. Finally, the magnetic dipole components of the CR are obtained from $m_x = (I_1+I_2)A$, $m_y = (I_3+I_4)A$, and $m_z = (I_5+I_6)A$. The resulting expression for the magnetic polarizability tensor is

$$\boldsymbol{\alpha}^{mm} = -j\omega A^2 \frac{2}{3}\left[\frac{1}{Z_{11}+Z_{12}-Z_{13}-Z_{14}}\begin{pmatrix} 2 & -1 & 1 \\ -1 & 2 & 1 \\ 1 & 1 & 2 \end{pmatrix} + \frac{1}{Z_{11}+Z_{12}+2Z_{13}+2Z_{14}}\begin{pmatrix} 1 & 1 & -1 \\ 1 & 1 & -1 \\ -1 & -1 & 1 \end{pmatrix}\right] \tag{14}$$

This magnetic polarizability tensor is anisotropic and exhibits two resonances, at those frequencies where $Z_{11} + Z_{12} - Z_{13} - Z_{14} \approx 0$ or $Z_{11} + Z_{12} + 2Z_{13} + 2Z_{14} \approx 0$. Only the even resonances of Table II appear in (14) because the excitation vector and the odd eigenvectors are orthogonal, i.e. $\mathbf{F}_m \cdot \mathbf{v}_i^{odd} = 0$ in (9). Just in the limit of no coupling between SRRs ($Z_{ij} = 0$ for $i \neq j$) both resonances converge to the single SRR resonance and $\alpha_{mm}$ become a scalar, as mentioned in Refs. [14-16]. However, it will be shown in the experimental section that this coupling cannot be neglected in most practical configurations.

Apart from the magnetic excitation studied above, it is well known that the Pendry's SRR can be excited by an external electric field [5, 6, 40]. Therefore, an electric response is also expected for this particular CR. The electric excitation of the rings on the cube is sketched in Fig. 5.

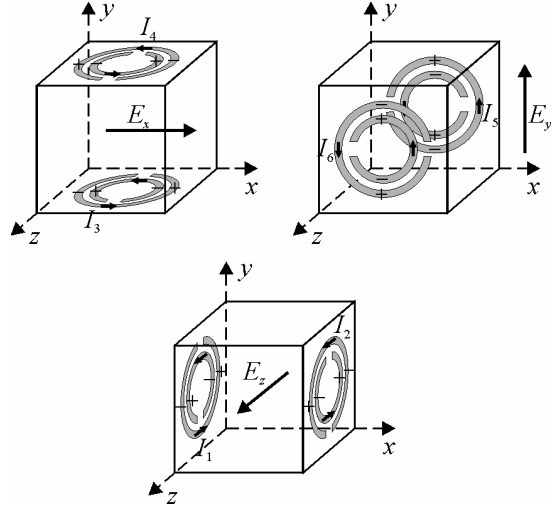

**Fig. 5:** Electric excitation of the cubic resonator made of Pendry's SRRs.

The external electric field $\mathbf{E}^{ext}$ can excite an SRR only if it has a non vanishing component in the plane of the particle, orthogonal to the imaginary line passing through the slits on the rings [5, 6, 40]. Therefore, only two SRRs are excited by each Cartesian component of $\mathbf{E}^{ext}$. Thus, for an external and homogeneous electric field of arbitrary direction, the excitation vector has the form

$$\mathbf{F}_e \propto (E_z^{ext},\ -E_z^{ext},\ -E_x^{ext},\ E_x^{ext},\ -E_y^{ext},\ E_y^{ext}). \tag{15}$$

Taking into account the sign of the charges induced over the rings, it is clear that the electric dipole has to be proportional to

$$\mathbf{p} \propto \begin{pmatrix} I_3 - I_4 \\ I_5 - I_6 \\ I_2 - I_1 \end{pmatrix} \tag{16}$$

The proportionality constants of (15) and (16) are given in the Appendix, equations (A2) and (A9). Finally, from equations (A9) of the Appendix, we get the following analytical formula for the electric polarizability tensor:

$$\boldsymbol{\alpha}^{ee} = \frac{32 d_{eff}^2}{3j\omega\pi^2} \left[ \frac{1}{Z_{11} - Z_{12} - Z_{13} + Z_{14}} \begin{pmatrix} -2 & 1 & -1 \\ 1 & -2 & -1 \\ -1 & -1 & -2 \end{pmatrix} + \frac{1}{Z_{11} - Z_{12} + 2Z_{13} - 2Z_{14}} \begin{pmatrix} -1 & -1 & 1 \\ -1 & -1 & 1 \\ 1 & 1 & -1 \end{pmatrix} \right], \tag{17}$$

where $d_{eff}$ is an effective distance between the metal strips on each SRR [5, 6]. Clearly, this electric polarizability tensor is anisotropic and exhibits two resonances, at those frequencies where $Z_{11} - Z_{12} - Z_{13} + Z_{14} \approx 0$ or $Z_{11} - Z_{12} + 2Z_{13} - 2Z_{14} \approx 0$. Only the odd resonances of Table II appear in (17), because the excitation vector and the even eigenvectors are orthogonal, i.e. $\mathbf{F}_e \cdot \mathbf{v}_i^{even} = 0$ in (9).

In summary, it has been shown that the coupling between the faces of the CR made of Pendry's SRRs shown in Fig. 1(a) splits the original resonance of a single SRR in four new resonances. Besides, both magnetic and electric polarizability tensors are anisotropic, as can be seen from equations (14) and (17). Finally, it will be worth to mention that the even modes of resonance can be also called magnetic modes, because they have a resonant magnetic moment and can be only excited by an external magnetic field, but not by an external electric field. Similarly, the odd modes of resonance are electric modes, because they present a resonant electric dipole which can be only excited by an external electric field. The reported conclusions are quite relevant for our analysis, because they show that a cubic arrangement of Pendry's SRRs will not be only anisotropic; it will also show several different resonances around the isolated SRR resonance, thus destroying any possibility of a single-resonance Lorentzian behavior of the metamaterial.

### b. Analysis of an isotropic cube

It was already shown in Sec. II(b) that, in order to ensure an isotropic behavior, the CR has to be invariant at least under the tetrahedron symmetry group $T = \{<\mathbf{1}, \mathbf{4}_x \cdot \mathbf{4}_y, \mathbf{4}_y \cdot \mathbf{4}_x>\}$. The $T$-CR shown in Fig. 1(c), made of six NB-SRRs [8, 13], is a good example of particle obeying this symmetry. By using the symmetry transformations and the rule (5), its impedance matrix can be significantly reduced to

$$\mathbf{Z} = \begin{bmatrix} Z_{11} & Z_{12} & Z_{13} & -Z_{13} & Z_{13} & -Z_{13} \\ Z_{12} & Z_{11} & -Z_{13} & Z_{13} & -Z_{13} & Z_{13} \\ Z_{13} & -Z_{13} & Z_{11} & Z_{12} & Z_{13} & -Z_{13} \\ -Z_{14} & Z_{13} & Z_{12} & Z_{11} & -Z_{13} & Z_{13} \\ Z_{13} & -Z_{13} & Z_{13} & -Z_{13} & Z_{11} & Z_{12} \\ -Z_{13} & Z_{13} & -Z_{13} & Z_{13} & Z_{12} & Z_{11} \end{bmatrix}, \quad (18)$$

where there are just three independent components. All eigenvalues and a complete set of orthonormal eigenvectors of this matrix are shown in Table III. If we compare Table III and Tab. II, we immediately find some similarities. In both cases the eigenvectors can be classified into even and odd types, and the eigenvalues in Table III can be obtained from those of Table II by making $Z_{14} = -Z_{13}$. Furthermore, the odd and even subspaces are kept, and only the two sub-spaces of the even eigenvectors are unified into a single sub-space due to the eigenvalue degeneration induced by the additional rotation symmetry $\mathbf{4}_x \cdot \mathbf{4}_y$.

Let us now analyze the resonances and polarizabilities of the $T$-CR by following the procedure developed in the previous subsection. By considering a homogeneous external magnetic excitation, the corresponding excitation vector (13) can only excite the even modes, thus leading to the isotropic magnetic polarizability tensor

$$\boldsymbol{\alpha}^{mm} = \frac{-2j\omega A^2}{Z_{11} + Z_{12}} \begin{bmatrix} 1 & 0 & 0 \\ 0 & 1 & 0 \\ 0 & 0 & 1 \end{bmatrix} \quad (19)$$

which, in fact, corresponds to the substitution $Z_{14} = -Z_{13}$ in (14). Furthermore, since only magnetic coupling between SRRs is considered in the frame of the present approximation, the self and mutual impedances are given by $Z_{11} = R + j\omega L + (j\omega C)^{-1}, Z_{12} = j\omega M_{12}$, with $R$, $L$, and $C$ being the resistance, self inductance, and self capacitance of each NB-SRR [8], and $M_{12}$ the mutual inductance between two NB-SRRs placed on opposite faces of the CR. Using these relations, the frequency of resonance of the CR can be calculated as $\omega_{res} = (C(M_{12} + L))^{-2}$, while the magnetic polarizability tensor takes the form

$$\boldsymbol{\alpha}^{mm} = \frac{2\omega^2 C A^2}{1 - \omega^2 (L + M_{12})C} \begin{bmatrix} 1 & 0 & 0 \\ 0 & 1 & 0 \\ 0 & 0 & 1 \end{bmatrix}. \tag{20}$$

This formula shows a Lorentzian-like magnetic response, similar to that of the single planar NB-SRR, but isotropic in 3D.

**Table III.** Eigenvalues and a complete set of orthonormal eigenvectors of the impedance matrix (5) corresponding to isotropic cubic resonators with symmetric under the tetrahedron group $T = \{<\mathbf{1}, \mathbf{4}_x \cdot \mathbf{4}_y, \mathbf{4}_y \cdot \mathbf{4}_x>\}$, as for instance the structure shown in Fig. 1(c).

|  | Eigenvalues, $z_i$ | Eigenvectors, $\mathbf{v}_i$ |
|---|---|---|
| Even modes (magnetic) | $Z_{11} + Z_{12}$ | $\frac{1}{\sqrt{2}}(0,0,1,1,0,0)$ $\frac{1}{\sqrt{2}}(0,0,0,0,1,1)$ $\frac{1}{\sqrt{2}}(1,1,0,0,0,0)$ |
| Odd modes (electric) | $Z_{11} - Z_{12} - 2Z_{13}$ | $\frac{1}{2\sqrt{3}}(1,-1,-2,2,1,-1)$ $\frac{1}{2}(1,-1,0,0,-1,1)$ |
|  | $Z_{11} - Z_{12} + 4Z_{13}$ | $\frac{1}{\sqrt{6}}(-1,1,-1,1,-1,1)$ |

With regard to the behavior of the considered CR under an external electric excitation, since the NB-SRRs can not be excited by an external electric field [8], the present theory predicts the absence of any resonant response to this excitation (of course, the CR will exhibit a non resonant electric polarizability due to the static electric moments induced on each ring, which is of no interest in the frame of the present discussion). However, experiments reported in [13] have shown that the considered CR exhibits a weak magneto-electric coupling with the same resonant frequency, $\omega_{res} = (C(M_{12} + L))^{-2}$, as in (20) Therefore, this effect does not affects nor the isotropy nor the single-resonance behavior of the considered CR. Since the tetrahedron symmetry group, $T$, does not include the inversion transformation, this result is not forbidden for $T$-CRs. Although the origin of this effect will be qualitatively explained bellow in the experimental section, it can be advanced that it is basically due to the capacitive coupling between SRRs, which is ignored in the equivalent circuit approximation developed in this section. Actually, the presence of this second order effect near the CR resonance shows how important the analysis of the spatial symmetries is in order to predict the behavior of metamaterial resonators: it seems that any effect not forbidden by symmetry will actually appear in practice, regardless of the equivalent circuit models. It is worth to recall here that this magneto-electric coupling disappears if the *T*-CRs are arranged in a *fcc* lattice with $b=2a$, as it was explained at the end of Sec. II(b).

## IV. EXPERIMENTS

For the experimental verification of the theory developed in the previous sections, some anisotropic and isotropic CRs were manufactured. Each CR was inserted into a standard WR430 waveguide as shown in Fig. 6. The testing procedure starts from the fact that the particle is isotropic if their polarizability tensors are invariant by any rotation. Therefore, all CRs were subjected to several rotations and the transmission coefficient through the waveguide was measured in a network analyzer HP-8510. If the measured CR were isotropic, then the measured transmission coefficient would remain invariant after rotations.

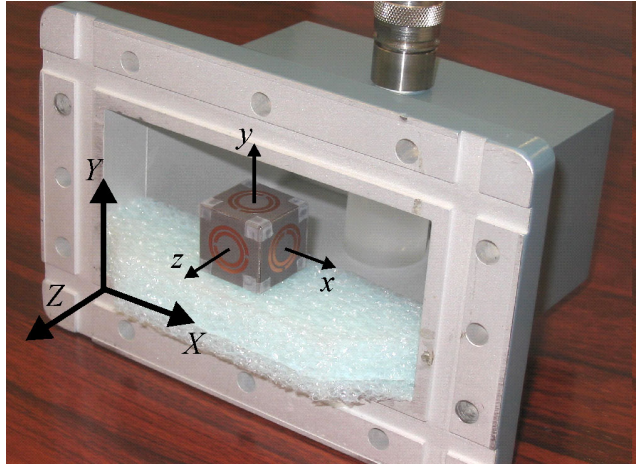

**Fig. 6:** (Color online) Experimental setup for checking the isotropy of cubic resonators. In the illustration a cubic resonator made of Pendry's SRR is placed inside a pair of standard waveguide-coaxial transitions WR430 connected to a network analyzer HP-8510-B. The transversal dimensions of the waveguide are 109 x 55 mm$^2$ and its frequency range is 1.7-2.6 GHz. The cube is hold by a piece of electromagnetically inert foam at an arbitrary orientation.

Namely, two anisotropic cubes made of Pendry's SRRs and Omega particles, and two isotropic cubes made of $C_2$-SRRs (actually NB-SRRs [8]) and $C_4$-SRRs (see insets in Fig. 7) were implemented. All SRRs were etched on ARLON 250-LX substrate with dielectric constant $\varepsilon_r = 2.43$, loss tangent tan $\delta < 0.002$, and thickness $t = 0.49$ mm. In order to check the similarity between the SRRs belonging to the same CR, their resonance frequencies were measured by placing each one in the E-plane of the waveguide, obtaining the following values: $f_0^{Pendry's\ SRR} = (2.321 \pm 0.002)$ GHz, $f_0^{\Omega} = (2.216 \pm 0.002)$ GHz, $f_0^{C4-SRR} = (2.399 \pm 0.001)$ GHz, $f_0^{NB-SRR} = (2.385 \pm 0.002)$ GHz. These results show that significant deviations from these values (of more than 0.002 GHz) in the measured resonances of the transmission coefficients for the CRs must be interpreted as a resonance splitting due to SRR couplings, and not due to fabrication imprecision. The SRRs were assembled over a cube of isotropic dielectric (ROHACELL 71 HF, $\varepsilon_r$ = 1.07, tan $\delta$ < 0.0002), of size $2\times2\times2$ cm$^3$.

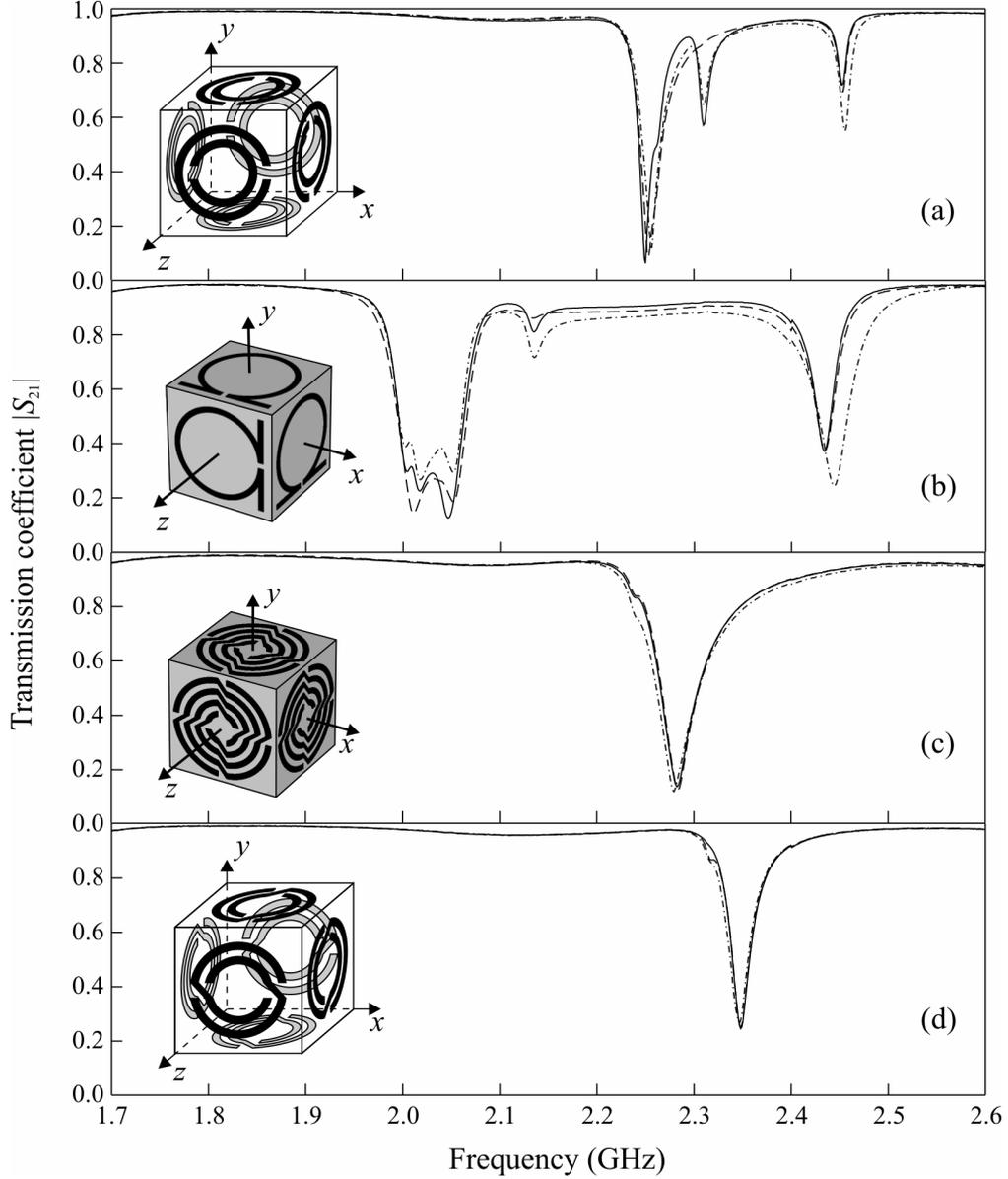

**Fig. 7:** Transmission coefficient ($|S_{21}|$) through a waveguide containing a cubic resonator made of Pendry's SRRs (a), Omega particles (b), $C_4$-SRRs (c), or $C_2$-SRRs (or NB-SRR) (d). Solid line: the particle is oriented with its axes ($x$, $y$, $z$) parallel to the waveguide axes ($X$, $Y$, $Z$) shown in Fig. 5. Dashed line: the first orientation is rotated by 45° along the $Y$-axis. Dash-dot line: the first orientation is rotated by 45° along the $Z$-axis and 45° along the $Y$-axis. The size of all cubes is $2\times2\times2$ cm$^3$. Dimensions of Pendry's SRRs: external radius $r_{ext}$ = 7 mm, width of the strip $w$ = 1.25 mm, distance between strips $d$ = 0.5 mm, and size of split gap $g$ = 1 mm. Dimensions of Ω: $r_{ext}$ = 8.5 mm, $w$ = 1 mm, $g$ = 1 mm, and the length of "legs" $l$ = 8 mm. Dimensions of $C_4$-SRR: $r_{ext}$ = 9.25 mm, $w$ = 1.25 mm, $d$ = 0.5 mm, and $g$ = 1.5 mm. Dimensions of $C_2$-SRR: external radius $r_{ext}$ = 7 mm, width of the strip $w$ = 1.25 mm, distance between strips $d$ = 0.5 mm, and size of split gap $g$ = 1 mm.

#### a. Anisotropic cubes

First, the CRs not satisfying the necessary spatial symmetries for isotropy were tested. Figure 7(a) shows the transmission coefficient through the waveguide loaded with the CR made of Pendry's SRRs. Two observations are apparent: there are three major resonance peaks and none of them stays invariant under rotations of the cube. Therefore this CR is anisotropic, as theoretically predicted by symmetry theory in Sec. II(b). However, the circuit model developed in Sec. III predicts the presence of four different resonances, not three, as it can be observed in Fig. 7(a). Although the fourth peak is not clearly visible in the figure, the lowest frequency peak splits into two peaks for other orientations not shown in Fig. 7, thus recovering the agreement with the theory.

To provide additional justification for the reported results, the cube composed of Omega particles was also experimentally tested. This cube has identical symmetry properties than the cube made of Pendry's SRR, i.e. it is invariant under inversion and the rotation $\mathbf{4}_y \cdot \mathbf{4}_x$. The transmission coefficient for this measurement is depicted in Fig. 7(b), where similar results as for the SRR cube can be observed. An important result of the reported measurements is that the relative frequency deviations between the different resonances of the SRRs and Omega CRs (a 10% or more with regard to the central frequency) is of the same order that the bandwidths reported for most SRR or Omega based negative-$\mu$ metamaterials. Therefore, as it was advanced in Sec. III, it can be guessed that any metamaterial made from such configurations will show multiple resonances inside the expected negative-$\mu$ frequency band.

#### b. Isotropic cubes

In order to shown the usefulness of spatial symmetries to provide isotropic resonators, the cubes made of $C_4$-SRRs and $C_2$-SRRs (see insets in Fig. 7(c,d)), satisfying the octahedron group $O$ and the tetrahedron group $T$, respectively, have been tested. As it was shown in Sec. II(b), both cubes are symmetric enough to be isotropic. The transmission coefficients for these CRs are shown in Fig. 7(c,d). It can be observed that the transmission does not depend on their orientations, thus demonstrating their isotropy. Besides, it can be seen that only one peak appear in both measurements, as it was predicted in Sec. III(b). It is worth noting that a similar result will be obtained for any CR satisfying any one of the five cubic symmetry point groups ($T$, $T_h$, $T_d$, $O$, and $O_h$).

The cubes analyzed in this section have not inversion symmetry and, as it was mentioned at the end of Sec. III, they could exhibit bi-isotropic behavior. However, from the experimental curves is impossible to see whether the analyzed CRs are bi-isotropic or not. To examine this possibility, electromagnetic simulations of a square waveguide loaded with $T$-CRs were made. The input port was fed by the $TE_{10}$ mode, while the $TE_{01}$ mode with orthogonal polarization was measured on the output port. The resulting cross-polarization transmission coefficient is shown in Fig. 8. The nonzero transmission means that the incident electric field can excite not only a parallel electric dipole, but also a parallel magnetic dipole. From reciprocity, it is also clear that an incident magnetic field can excite both magnetic and electric dipoles parallel to the exciting field. This result clearly shows the bi-isotropic behavior of the $C_2$-SRR cube. In order to show that the bi-isotropy can be avoided by including the inversion symmetry in the configuration, a similar simulation was carried out for the $T_h$-CR shown in Fig. 1(d), already proposed in [13], which possesses inversion symmetry. The transmission coefficient, not depicted here, was almost zero and of the same order as the transmission through the waveguide without any resonator, thus showing that this last configuration is not bi-isotropic.

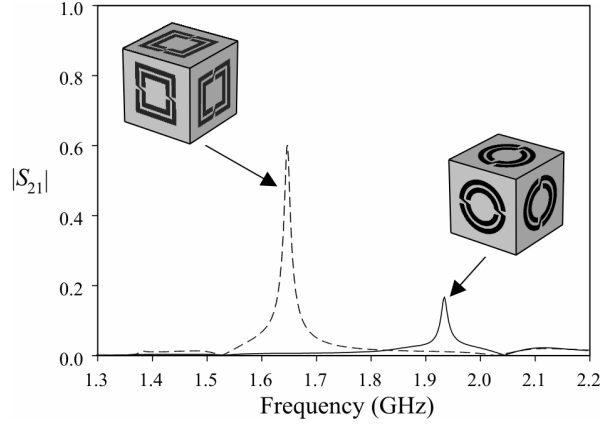

**Fig. 8:** Simulated transmission coefficient ($|S_{21}|$) through the cross polarization waveguide setup filled by *T*-CRs made of circular (solid line) and square NB-SRRs (dashed line). Dimensions of the circular NB-SRR: external diameter $2r_{ext}$ = 20 mm, width of the strip $w$ = 2 mm, distance between strips $d$ = 1 mm, and size of split gap $g$ = 1.6 mm. The square NB-SRRs has similar dimensions and the same external perimeter. Cube edge was 24 mm long.

As it was already reported, the bi-isotropy of the *T*-CR cannot be explained by the circuit model proposed in Sec. III. The explanation of this effect seems to rely on the capacitive coupling between the edges of two SRR on adjacent faces of the cube. To understand this in a qualitative way, let us assume that the cube is driven by an external magnetic field feeding only two resonators in the cube, as it is shown in Fig. 9.

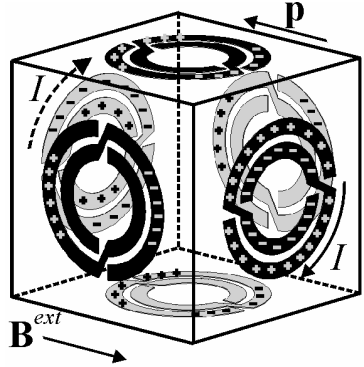

**Fig. 9:** Illustration of the bi-isotropic behavior of a *T*-CR made of NB-SRRs driven by an external magnetic field.

This figure also depicts the current and corresponding charges induced by the exciting field on each NB-SRR. Due to the inversion symmetry of the NB-SRR, the electric dipole generated by the excited resonators is zero [8]. However, Fig. 9 also shows how the induced resonant charges polarize the other (not excited) rings. These are polarized in such a way that the CR acquires a net electric dipole, as it is sketched in the figure. According to the above explanation, it is expected that magnetoelectric coupling will increase if the capacitance between the edges of neighboring resonators grows. To check this hypothesis the transmission of the cross-polarized wave when square rings are used (dashed line in Fig. 8 ) was also simulated. The enhancement of the magneto-electric coupling can be clearly observed in this case.

At the end of Sec. II it was mentioned that this bi-isotropic behavior would disappear in an *fcc* cubic lattice with *b = 2a* (see Fig. 3). In that Section this behavior was predicted on the basis of the particular symmetry of this specific lattice. The illustration in Fig. 9 of the bi-isotropy of an isolated CR made of six identical NB-SRRs, also provides a qualitative physical interpretation of such result: if additional SRRs were added to Fig. 9 in order to make an *fcc* cubic lattice, it becomes apparent that the induced charges on the non-resonant additional SRRs will cancel the total electric dipole shown in the figure.

## V. CONCLUSIONS

A systematic approach to the design of isotropic magnetic metamaterials by using isotropic cubic magnetic resonators in a cubic lattice has been developed. For this purpose cubic magnetic resonators obeying some cubic point groups of symmetry ($T$, $T_d$, $T_h$, $O$, or $O_h$) placed in a cubic Bravais' lattice (*sc*, *bcc*, or *fcc*) were analyzed. Special care has been taken in the study of the symmetry of the constitutive elements (also called cubic resonators or CRs). For practical reasons, CRs made of six modified SRRs assembled over the surface of a cube were considered. The connection between the orientations of these SRRs over the cube and the cubic point groups of symmetry has been analyzed. Starting from this analysis, some particular examples of non-isotropic and isotropic CRs were analyzed, manufactured and measured. It was analytically and experimentally shown that the lack of the necessary symmetry leads to an anisotropic response. In experiments, the transmission through a waveguide loaded with the manufactured CRs was measured, getting a strong dependence of this parameter on the orientation for anisotropic CRs, while transmission was invariant with respect to the orientation for isotropic CRs. Furthermore, the splitting of the isolated SRR resonances into several resonances was observed in anisotropic CRs. This effect is absent in isotropic CRs, which always show a single resonance. Most of these effects were theoretically explained by using an equivalent circuit model which takes into account the magnetic couplings between the SRRs making the analyzed CRs.

From a practical standpoint, we have found that using some low symmetry CRs, pertaining to the tetrahedral groups $T$ or $T_h$, placed in a cubic Bravais lattice is enough to provide isotropy in three dimensions. Using CRs with lower symmetry results in an anisotropic behavior, even if the dipole representation of the SRRs suggests an isotropic behavior. In general, using cubic resonators pertaining to a symmetry group which does not include inversion (such as the symmetry group $T$) produce a bi-isotropic behavior, even if the isolated SRRs making the metamaterial do not present magneto-electric coupling. However, this bi-isotropy can be avoided by a proper choice of the lattice. In particular it has been shown that cubic resonators pertaining to the aforementioned $T$ group placed in an *fcc* lattice with the appropriate periodicity can produce a purely magnetic isotropic behavior.

We hope that the reported results will pave the way to the design of isotropic three-dimensional periodic metamaterials with a resonant magnetic response, including negative permeability and left-handed metamaterials.

**Acknowledgements**

This work has been supported by the Spanish Ministry of Education and Science under project contract TEC2004-04249-C02-02, and by the Grant Agency of Czech Republic under project 102/03/0449. Authors also thank to Esperanza Rubio for manufacturing the resonators used in the experiments.

**APPENDIX: ELECTRIC EXCITATION OF THE CR MADE OF PENDRY'S SRR**

Let us assume a single SRR placed in the *xy*–plane with its two slits along the *x*-axis. The electromotive force can be approximated by averaging the path integral of the external electric field through the gap along the circumference of the particle, so that

$$fem = \langle \mathbf{E}^{ext} \cdot \mathbf{d}_{eff}(\varphi) \rangle = 2 E_y^{ext} d_{eff} \frac{1}{\pi} \int_0^\pi \cos(\varphi - \pi/2) \, d\varphi = \frac{4}{\pi} d_{eff} E_y^{ext} . \tag{A1}$$

It is worth to note that the two behalves of the SRR are polarized in the same direction [5, 6], so that it justify the factor 2 in front of the integral and its integration domain $(0, \pi)$. Now, let us generalize the electromotive force of (A1) to get the "excitation vector" for the CR made of Pendry's SRR shown in Fig. 1(a). B. Taking into account the sketch of the excitation shown in Fig. 5, it easy to get the following electric excitation vector:

$$\mathbf{F}_e = \frac{4}{\pi} d_{eff} (E_z^{ext}, \; -E_z^{ext}, \; -E_x^{ext}, \; E_x^{ext}, \; -E_y^{ext}, \; E_y^{ext}) . \tag{A2}$$

In what follows, for simplicity, the superscript *ext* will be avoided. By introducing (A2) in equation (9) we get the associated currents:

$$\begin{pmatrix} I_1 \\ I_2 \\ I_3 \\ I_4 \\ I_5 \\ I_6 \end{pmatrix} = \frac{\frac{4}{3\pi} d_{eff}}{Z_{11} - Z_{12} - Z_{13} + Z_{14}} \begin{pmatrix} E_x + E_y + 2E_z \\ -E_x - E_y - 2E_z \\ -2E_x + E_y - E_z \\ 2E_x - E_y + E_z \\ E_x - 2E_y - E_z \\ -E_x + 2E_y + E_z \end{pmatrix} + \frac{\frac{4}{3\pi} d_{eff}}{Z_{11} - Z_{12} + 2Z_{13} - 2Z_{14}} \begin{pmatrix} -E_x - E_y + E_z \\ E_x + E_y - E_z \\ -E_x - E_y + E_z \\ E_x + E_y - E_z \\ -E_x - E_y + E_z \\ E_x + E_y - E_z \end{pmatrix} . \tag{A3}$$

The electric dipole for a single SRR can be expressed in terms of a linear charge density $\lambda$ as [5, 6] as

$$p_y = 2 \lambda r_0 d_{eff} \int_0^\pi \cos(\varphi - \pi/2) \, d\varphi = 4 \lambda r_0 d_{eff} , \tag{A4}$$

where $d_{eff}$ is an effective distance between the two metallic strips forming the SRR. The charge density on the inner, $I_i$, and outer rings, $I_o$, of the SRR can be calculated by means of the conservation law as it follows:

$$\frac{dI_{i,o}}{d\theta} = j\omega r \lambda_{i,0} \; \Rightarrow \; \lambda_{i,0} = \frac{1}{j\omega r} \frac{dI_{i,o}}{d\theta} . \tag{A5}$$

Since the SRR size is much smaller than one wavelength, we can suppose a linear variation of $I_{i,o}$ respect to the angle $\phi$, taking maximum value $I$ at the center of the metal strip and zero at its ends, as in Refs. [4, 5]. Then

$$|\lambda| = \frac{1}{j\omega r} \frac{|I|}{\pi} . \tag{A6}$$

Although $I_i$ and $I_o$ are not uniform through $\phi$, the sum of both, $I_i + I_o$, is approximately constant and equal to the current $I$, which is actually the effective current associated to the averaged loop. By equations (A4) and (A6), we obtain

$$\left|p_y\right| = \frac{4d_{eff}}{j\omega\pi}|I|. \tag{A7}$$

Now, we can calculate the total electric moment of the SRR cube by adding the six moments. By considering the equation (A7) and taking into account the signs of the charges shown in Fig. 5, we obtain the electric dipole

$$\mathbf{p} = \frac{4d_{eff}}{j\omega\pi}\begin{pmatrix} I_3 - I_4 \\ I_5 - I_6 \\ I_2 - I_1 \end{pmatrix} \tag{A8}$$

Finally, by substituting the currents of equation (A3) into (A8), we get de electric dipole in terms of the components of the external electric field:

$$\mathbf{p} = \frac{32d_{eff}^2}{3j\omega\pi^2}\left[\frac{1}{Z_{11} - Z_{12} - Z_{13} + Z_{14}}\begin{pmatrix} -2E_x + E_y - E_z \\ E_x - 2E_y - E_z \\ -E_x - E_y - 2E_z \end{pmatrix} + \frac{1}{Z_{11} - Z_{12} + 2Z_{13} - 2Z_{14}}\begin{pmatrix} -E_x - E_y + E_z \\ -E_x - E_y + E_z \\ E_x + E_y - E_z \end{pmatrix}\right]. \tag{A9}$$

**Referentes**